\documentclass[useAMS,usenatbib]{mn2e}

\usepackage{epsfig}
\usepackage{graphicx}
\usepackage{ulem}
\usepackage{color}
\usepackage{amsmath}
\usepackage{multirow}
\usepackage{amssymb}
\def\msun{M_\odot}
\def\Dd{D_{\rm L}}
\def\Ds{D_{\rm S}}
\def\mearth{M_\oplus}

\title[Free-floating Planets: Microlensing Predictions]
      {Free-floating planets from core accretion theory: microlensing predictions}
\author[Ma et al.]{Sizheng Ma$^{1}$, Shude Mao$^{1,2,3}$\thanks{smao@tsinghua.edu.cn}, Shigeru Ida$^{4}$, Wei Zhu$^{5}$, Douglas N.C. Lin$^{1,2,6}$\\
$^{1}$ Department of Physics and Center for Astrophysics, Tsinghua University, Haidian District, Beijing 100084, China\\
$^{2}$ National Astronomical Observatories, 20A Datun Road, Chinese Academy of
Sciences, Beijing, 100012, China \\
$^{3}$ Jodrell Bank Centre for Astrophysics, School of Physics and Astronomy, 
University of Manchester, Alan Turing Building, \\
~~Oxford Road, Manchester M13 9PL, UK  \\
$^{4}$ Earth-Life Science Institute, Tokyo Institute of Technology, Okayama, Meguro-ku, Tokyo 152-8550, Japan \\
$^{5}$ Department of Astronomy, The Ohio State University, 140 W. 18th Avenue, Columbus, OH 43210, USA\\
$^{6}$ University of California Observatories, Lick Observatory, University of California, Santa Cruz, CA 95064\\
}
\begin{document}
\include{journaldefs}
\date{Accepted ...... Received ...... ; in original form......   }

\maketitle
\label{firstpage}

\begin{abstract}
    We calculate the microlensing event rate and typical time-scales for the free-floating planet (FFP) population that is predicted by the core accretion theory of planet formation. The event rate is found to be $\sim 1.8\times 10^{-3}$ of that for the stellar population. While the stellar microlensing event time-scale peaks at around 20 days, the median time-scale for  FFP events ($\sim$0.1 day) is much shorter. Our values for the event rate and the median time-scale are significantly smaller than those required to explain the \cite{Sum+11} result, by factors of $\sim$13 and $\sim$16, respectively. The inclusion of planets at wide separations does not change the results significantly.
    This discrepancy may be too significant for standard versions of both the core accretion theory and the gravitational instability model to explain satisfactorily. Therefore, either a modification to the planet formation theory is required, or other explanations to the excess of short-time-scale microlensing events are needed. Our predictions can be tested by ongoing microlensing experiment such as KMTNet, and by future satellite missions such as WFIRST and Euclid.

\end{abstract}
\maketitle


\section{Introduction}
\label{sec:intro}

Microlensing as a method to detect planets around a host star was
proposed 25 years ago (\citealt{MP91, GL92}). So far more than 40
planets discovered by microlensing have been published\footnote{http://exoplanet.eu}, with many more
discovered but yet to be published. Microlensing probes the
planet population beyond the snow line in the planet mass vs. host star separation
parameter space,  and is complementary to other detection methods (see \citealt{Gau12, Mao12} for reviews).

A significant advantage of the microlensing method is that it can detect free-floating planets (FFPs). Events caused by FFPs have short time-scales because time-scales scale as $M^{1/2}$, where $M$ is the lens mass (cf. eq. \ref{eq:tE}). For example, the typical time-scale is of order 1 day for a Jupiter-mass FFP and a few hours for an Earth-mass FFP.
Ground-based high-cadence surveys, such as the Microlensing Observations in Astrophysics (MOA-II; \citealt{Bon+01, Sum+03}), the Optical Gravitational Lensing Experiment (OGLE-IV; \citealt{USS15}), and now the Korean Microlensing Telescope Network (KMTNet, \citealt{Kim+16,Hen+14}), are in principle able to detect such events.

The MOA-II collaboration in fact detected 10 events with an Einstein radius crossing time-scale $t_{\rm E} <$\,2 days among 474 well characterized events in their 2006-2007 data set.
After correcting the detection efficiency, \cite{Sum+11} concluded that a population of unbound or distant Jupiter-mass objects that are almost twice as common as stellar objects is required to explain the observed time-scale distribution.


An immediate question is whether the result in \citet{Sum+11} is consistent with predictions from planet formation models. The two competing theories of planet formation are core accretion (e.g., \citealt{IL04, M+09,ILN13}) and gravitational instability (e.g., \citealt{Bos06}). In this paper, we focus on the former, but also address the influence of the latter in Section 6. Population synthesis models in the core accretion regime have taken into account many physical processes (such as migration and collision), and can provide detailed predictions about the properties of not only the bound planet population (\citealt{IL04, Mor+09}) but also FFPs. These FFPs are formed inside the proto-planetary disk and ejected due to dynamical interactions with other objects (see also \citealt{Pfy+15}).

The purpose of this work is to study the microlensing signature of the FFP population predicted by the popular core accretion theory. The structure of this paper is as follows: in \S2, we review the properties of the FFP population as predicted by \cite{ILN13}; in \S3 and \S4, we describe microlensing basics and the simple Galactic model that we use for creating microlensing observables from our FFP populations; our results are presented in \S5; and finally in \S6, we summarise and discuss our work.

\section{Properties of the FFP Populations}
\label{sec:stats}

\begin{figure}
\includegraphics[width=\hsize]{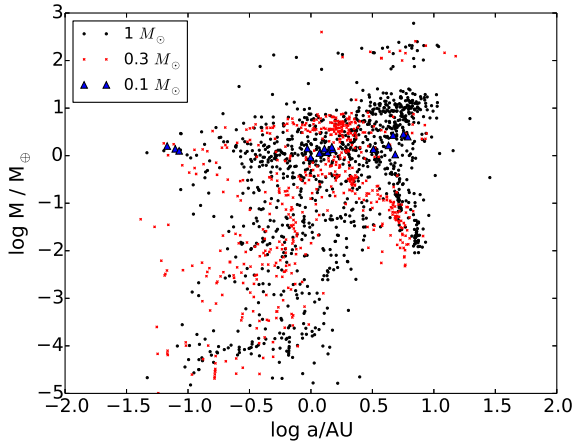}
\caption{The mass and semi-major axis (before ejection) distributions of the FFP populations used in this work. The three different symbols are used for FFPs ejected from three different host star masses (1, 0.3, and 0.1 $msun$). In total 1000 planetary systems have been simulated, with 1069, 571 and 17 ejected planets, respectively.}
\label{fig:m-a}
\end{figure}

\begin{figure}
\includegraphics[width=\hsize]{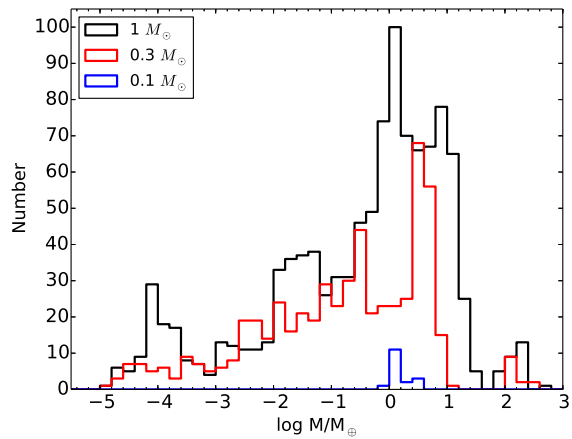}
\caption{The mass function of the free-floating planets. The three lines are for three host star masses, i.e., $1, 0.3$ and $0.1M_\odot$ respectively.}
\label{fig:m}
\end{figure}

We adopt the planet population synthesis model that is described in \citet{ILN13}. Note that their simulations have different parameter settings, for example regarding the orbital migration of planets, which may lead to slight differences in the ejected planets' population.
In this work, we use the results from their most up-dated simulations, in which the planet migrations are described as non-isothermal processes. Readers can find the distributions of the bound planet population predicted by such simulations in Fig.~14 of \cite{ILN13}.  Their calculation starts from very small planetary seeds; planets with mass larger than $1\times 10^{-5}\mearth$ have already increased their masses significantly and are thus not affected by initial conditions.

\begin{table}
    \centering
    \caption{Summary of the FFP populations used in this work. The last
      column indicates the fraction of stellar systems with ejected planets; the third column indicates the mean and median masses of individual ejected planets for each stellar mass; the fourth column indicates those of the total planetary mass ejected from individual sytem. All the planetary masses are in units of $M_\oplus$.}
    \begin{tabular}{|c|c|c|c|c|} \hline
        $M_\star$ &    & Individual FFP  & Total  FFP & Fraction  \\
        &    &mass ($M_\oplus$) & mass ($M_\oplus$)  &  \\ \hline
  \multirow{2}{*}{1 M$_{\odot}$}   
                  & mean      &7.2 & 44.2     &                  
                                                                      \multirow{2}{*}{17.5 $\%$} \\
                  & median    &0.8               &16.8                        \\    
                  \hline
     \multirow{2}{*}{0.3 M$_{\odot}$}     
                  & mean      &5.2              & 24.5              &            
                                                                       \multirow{2}{*}{12 $\%$} \\
                  & median    &0.3              &5.1                       \\ 
                  \hline
      \multirow{2}{*}{0.1 M$_{\odot}$}  
                  & mean      &1.6              &3.3                &       
                                                                      \multirow{2}{*}{0.8 $\%$} \\
                  & median    &1.4               &3.0                 &       \\ 
                  \hline
    \end{tabular}
\label{tab:planet-stats}
\end{table}

\begin{figure}
\includegraphics[width=\hsize]{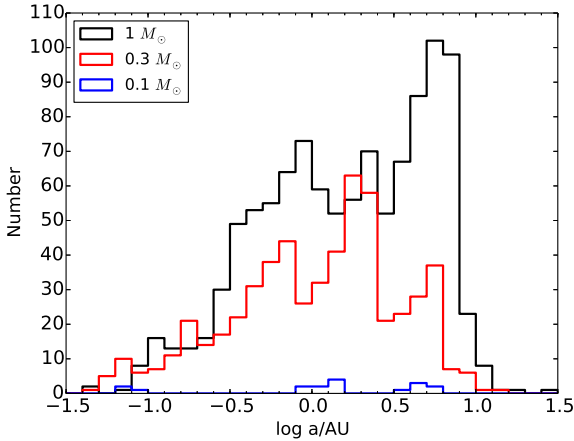}
\caption{The distribution of the semi-major axis just before the ejection. The symbols are the same as in Fig. \ref{fig:m}.}
\label{fig:a}
\end{figure}

Their results for three host star masses, $1 \msun, 0.3 \msun$ and $0.1\msun$, are used here. For each stellar mass, 1000 systems are simulated. In total there are respectively 1069, 571 and 17 ejected planets.
Fig.~\ref{fig:m-a} shows the ejected planets in the mass vs. semi-major axis plane just before ejection while Figs.~\ref{fig:m} and \ref{fig:a} show the histograms of the masses and semi-major axes, respectively. For $0.3 \msun$ and $1\msun$ stars, one sees an ejected population of Jupiter-mass planets and then more abundant super-Earth planets, followed by an extended tail down to $10^{-5} \mearth$.

Only a minority of simulated stellar systems produce free-floating planets: for $1, 0.3$ and $0.1\msun$, these fractions are 17.5\%, 12\% and 0.8\%, respectively (see Table \ref{tab:planet-stats}). The third column of Table~\ref{tab:planet-stats} shows the mean and median masses of all ejected planets. In each case, the median planet mass is much smaller than the mean due to the extended tail down to very low masses. It is interesting to note that the mean values are only a few Earth masses in all cases.

\begin{figure}
\includegraphics[width=\hsize]{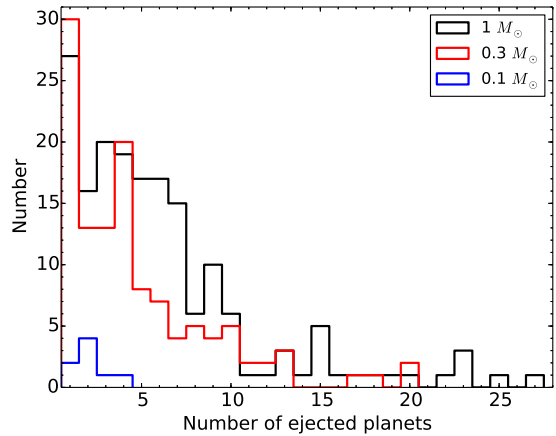}
\caption{Histogram of the numbers of ejected planets from individual systems.}
\label{fig:n}
\end{figure}

For systems that eject planets, each system tends to eject more than one planet. This is illustrated in Fig. \ref{fig:n}. For $1, 0.3$ and $0.1\msun$ stars, each system on average ejects $6.1, 4.8$, and $2.1$ planets, respectively. We sum up their total masses and list the mean and median values in the fourth column of Table~\ref{tab:planet-stats}; the mean values of total ejected mass are $24.5\mearth$ and $44.2\mearth$ for $0.3\msun$ and $1\msun$ stars respectively. For $0.1\msun$ stars, the value is significantly smaller ($3.3\mearth$).

The ejected planets are generally scattered by emerging gas giants which have sufficient mass to induce recoil speeds comparable to or larger than their escape velocities.  Since gas giants form less frequently around low-mass stars, the ejected fraction from their domain is much lower than that for the FGK stars.  Around 1 $M_\odot$ stars, the initial location of the escapers is preferentially outside 1AU (see Figs.\,\ref{fig:results}-\ref{fig:a}) because 1) the binding energy by the host stars decreases as the separation increases and 2) birth places of the gas giants are preferentially near the snow line.  The timescale for forming progenitor cores increases with the semi-major axis. Beyond 10 AU from the host star, the formation probability of gas giant declines.  Nonetheless, a few Neptune-mass failed cores may still scatter some residual planetesimals out of the gravitational confines of their host stars.

\section{Microlensing Basics}
\label{sec:microlensing}

\subsection{Event time-scale}

The Einstein radius ($r_{\rm E}$) crossing time-scale is given by 
\begin{equation}
\begin{split}
t_{E} \equiv \frac{r_{\rm E}}{v_t}\approx &19~{\rm d}~\sqrt{4\times\frac{\Dd}{\Ds}\left(1-\frac{\Dd}{\Ds}\right)}\left(\frac{\Ds}{8~{\rm kpc}}\right)^{1/2}  \\
&\times\left(\frac{M}{0.3 \msun}\right)^{1/2}\left(\frac{v_{t}}{200~{\rm km~s^{-1}}}\right)^{-1},
\end{split}
\label{eq:tE}
\end{equation}
where $\Ds$ and $\Dd$ are the distances to the source and the lens, respectively, $M$ is the lens mass and $v_{t}$ is the transverse velocity (\citealt{Mao12}). The time-scale is proportional to $M^{1/2}$; for a $0.3 \msun$ star, it is around 20 days.

\subsection{Event rate and optical depth}

The optical depth is the probability that a given source falls into the Einstein radius of any lensing star along the line of sight, and is defined as (see, e.g., \citealt{Pac96} for reviews)
\begin{equation}
\tau=\int_{0}^{\Ds} n(\Dd)\,(\pi r_{E}^{2})\,d\Dd,
\end{equation}
where $n(\Dd)$ is the number density of lenses. For a source population distributed at various distances, it can be written as  (e.g., \citealt{KP94, WM05})
\begin{equation}
\begin{split}
\langle \tau \rangle_{\gamma} = &\frac{4\pi G}{c^{2}} \int_{0}^{\infty} d\Ds\Ds^{2-\gamma}\rho(\Ds)\int_{0}^{\Ds}d\Dd\rho(\Dd)\Dd\frac{\Ds-\Dd}{\Ds}  \\
&\times \left[\int_{0}^{\infty} d\Ds\Ds^{2-\gamma}\rho(\Ds) \right]^{-1},
\end{split}
\end{equation}
where $\rho(\Ds)$ and $\rho(\Dd)$ are the densities of the source and lens at each position, and $\gamma$ is related to the source luminosity function (we take $\gamma=0$ as in \citealt{WM05}).

The event rate ($\Gamma$) describes the number of microlensing events per unit time for a given number of monitored stars ($N$), and is given by
\begin{equation}
\Gamma=\frac{2N}{\pi}\int\frac{d\tau}{t_{E}},
\end{equation}
which can be calculated as
\begin{equation}
\begin{split}
\Gamma=&\frac{4G^{\frac{1}{2}}}{c}\int_{0}^{\infty} d\Ds\Ds^{2-\gamma}\rho(\Ds) \\
&\times\frac{\int_{0}^{\Ds}d\Dd\rho(\Dd)v[\Dd(\Ds-\Dd)/M\Ds]^{1/2}}{\int_{0}^{\infty} d\Ds\Ds^{2-\gamma}\rho(\Ds)}.
\end{split}
\end{equation}

\section{Galactic Model} \label{sec:model}

Given the uncertainties in the planet population synthesis model (see \citealt{ILN13}), it is sufficient to adopt a simple Galactic model that describes the stellar density and kinematics. We therefore adopt the model used in \cite{WM05} (see also, e.g., \citealt{HG96, AKR15}).

\subsection{Lens and source density distributions}

We use the G2 (barred) model \citep{Dwe+95}, with $R_{\rm max}=5\,$kpc. The distance to the Galactic Centre is taken to be 8\,kpc. This is different from the 8.5\,kpc value in \cite{Dwe+95}, so other lengths are scaled down proportionally. The major axis of the bar is inclined by $30^{\circ}$ with respect to the sight-line toward the Galactic Centre, as suggested by several recent studies (\citealt{Cao+13, WG13}), instead of $13.4^\circ$ as adopted in \citealt{WM05}.

The model requires normalisation to observed star counts and we use HST star counts as in \cite{WM05}. In addition, we assume that the stellar mass function is a $\delta$-function at (1, 0.3, 0.1) $\msun$. Correspondingly, each star on average ejects (1.069, 0.571, 0.017) planets, with masses randomly drawn from the distributions shown in Fig.~\ref{fig:m-a}. We present the results toward the direction $(l, b)=(1^\circ.50,-2^\circ.68)$ as in \cite{WM05}; for other lines of sight, the ratio between the stellar  and planetary microlensing events is nearly unchanged.

\subsection{Kinematics}

We assume that the observer follows the Galactic rotation, so the velocities in the longitudinal and latitudinal directions are given by
\begin{equation}
    v_{{\rm O},l}=220\,{\rm km\,s^{-1}},   ~~~~~~~ v_{{\rm O},b}=0\, {\rm km\,s^{-1}}.
\end{equation}
For rotation of the disc, we use $v_{\rm rot}=220$ km/s; for the bar, the $v_{\rm rot}$ is taken to be (\citealt{HG96})
\begin{equation}
\begin{split}
&v_{\rm rot}=v_{\rm max}\left(\frac{x}{\rm 1\,kpc}\right),~~~~~R < 1\,{\rm kpc},  \\
&v_{\rm rot}=v_{\rm max}\left(\frac{x}{R}\right),~~~~~R \geqslant 1\,{\rm kpc},
\end{split}
\end{equation}
 where $R=(x^{2}+y^{2})^{1/2}$ and $v_{\rm max}=100\,{\rm km\,s^{-1}}$. The coordinates $(x,y,z)$ have their origin at the Galactic Centre, and the $x$ and $z$ axes point towards the Earth and the North Galactic Pole respectively.
In addition to the systematic rotations, stars also have random motions (assumed to be Gaussian). For the disc, the velocity dispersions are taken to be $\sigma_{l,b}=(30, 20)\,{\rm km\,s^{-1}}$, and for the bar, $\sigma_{x,y,z}=(110,82.5,66.3) \,{\rm km\,s^{-1}}$ along the major, intermediate and minor (vertical) axes. These numbers are also the same as \cite{HG96}. The total velocity for a disc or bar star is a sum of the rotation and local random motions:
\begin{equation}
v_{l}=v_{\rm rot}+v_{\rm rand,l}  \qquad  v_{b}=v_{\rm rand,b}.
\label{eq:velocity}
\end{equation}
The transverse velocity, decomposed into two directions, $v_{l}$ and $v_{b}$, is
\begin{equation}
v_{l,b}= \left[(v_{\rm L}-v_{\rm O})+(v_{\rm O}-v_{\rm S}) \frac{\Dd}{\Ds} \right]_{l,b},
\end{equation}
where
$v_{\rm L}$ and $v_{\rm S}$ are the lens and source velocities drawn from the distribution given by eq.\,(\ref{eq:velocity}). The total transverse velocity is
\begin{equation}
v_t=(v_{l}^{2}+v_{b}^{2})^{1/2}.
\end{equation}

\section{Results}

\begin{figure}
\includegraphics[width=\hsize]{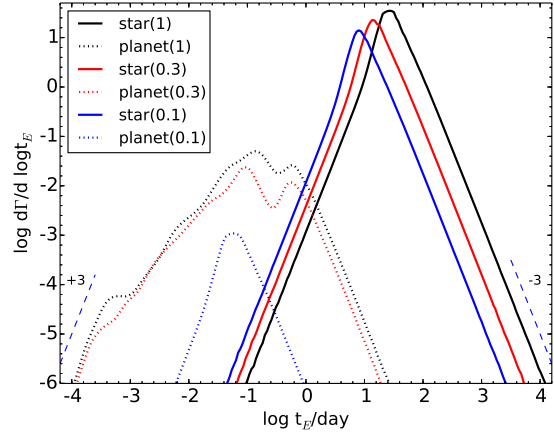}
\caption{The event rate as a function of timescale $t_E$; both axes are on
  logarithmic scales. The three solid curves are for the events produced
by three stellar masses, $1, 0.3$ and $0.1\msun$, respectively. The
three corresponding planetary curves (dotted lines) are shown on the left. The
asymptotic slopes \citep{MP96} at short and long time-scales are indicated by two
dashed lines.}
\label{fig:results}
\end{figure}

Our main results are illustrated in Fig.~\ref{fig:results} and also summarized in Table~\ref{tab:results30}.  The three curves on
the right are the predicted event rates for $1, 0.3$ and 0.1$\msun$
stars. Both the time-scale and the event rate scale as $M^{1/2}$
(a linear shift on the log-log plot). The event rate is about 14.0,
7.7 and 4.4 per million stars per year (see also \citealt{HG96, WM05, AKR15}).


The corresponding curves for the free-floating populations are shown on
the left. We notice that for the $0.3$ and $1\msun$ curves, there are two distinct peaks at around $0.1$\, and 1 day, respectively, these reflect
the super-Earth and the Jupiter-mass populations shown on
Fig.~\ref{fig:m-a}; the extended tails are primarily due to very low mass
planets down to $10^{-5} M_\oplus$. As expected, all the curves eventually follow the
asymptotic behaviours at the very short and very long time-scales \citep{MP96}. 

Since typical lenses are predicted to be around 0.3$\msun$, we compare the
predicted distributions for this mass with the results of
\cite{Sum+11}. The event rate for FFPs is found to be $\sim 1.8\times 10^{-3}$ times that for the stellar population and the median time-scale is shorter by a factor of $\sim 16$ than that for stars. If we focus on the FFPs,  we find their median time-scale is much shorter $\sim 0.1$ days than the peak of
$\sim 1.6$ days in \cite{Sum+11} (see their Fig.\,2). In addition, the
predicted event rate for FFPs is about $\sim 13$ times lower than those found in the MOA
data. 
Thus there is a serious discrepancy between observations and
theory, a point we will return to in the next section.

\begin{table}
    \centering
    \caption{The predicted optical depth $\tau$, event rate  (per million stars per year), $\Gamma$, and the mean and median (last column) time-scales in days for the stars and planets respectively. The bar angle is $30^{\circ}$.}

    \begin{tabular}{|p{0.8cm}|c|c|c|p{0.8cm}|p{0.8cm}|} \hline
    \multicolumn{2}{|c|}{}    & $\tau$ & $\Gamma$   & $\left<\log(t_{E})\right>$ & $\log t_{E, {\rm med}}$\\ 
    \hline
     \multirow{2}{*}{0.1 $M_{\odot}$}   
                            & planet      & $1.4\times 10^{-13}$   &$5.0 \times 10^{-4}$ &-1.2  &-1.3\\
                            & star        & $1.8\times 10^{-7}$     &4.4                 & 0.9  &0.9\\ 
                               \hline
     \multirow{2}{*}{0.3 $M_{\odot}$}     
                            & planet      & $1.2 \times 10^{-11}$ &$1.4 \times 10^{-2}$ & -1.0 &-1.0\\ 
                            & star        & $5.4 \times 10^{-7}$  &7.7                   &1.2  &1.1 \\ 
                              \hline
      \multirow{2}{*}{1 $M_{\odot}$}  
                            & planet      & $4.4 \times 10^{-11}$ &$3.9 \times 10^{-2}$ &  -0.9 &-0.9\\ 
                            & star        & $1.8 \times 10^{-6}$  &14.0                 &  1.4  &1.4\\  
                            \hline
    \end{tabular}
\label{tab:results30}
\end{table}

\section{Summary and Discussion}
\label{sec:discussion}

In this paper, we have used the FFP population in the core accretion
theory of \cite{ILN13} to make predictions for the microlensing event rate and typical time-scale. This is accomplished with a simple Galactic model in terms of density and kinematics. This model has limitations, for example the mass function is not realistic. As a result, the time-scale distribution is somewhat narrower than the observed one, and our median time-scale (12.6 days) for $0.3M_\odot$ is shorter than the one with more realistic mass functions (19.2 days, \citealt{WM05}).


We have to emphasise that the FFPs we used here are completely ejected from their systems. However, the microlensing effects of planets on wide-separation orbits may be similar to FFPs.  We therefore check planets that are away from their host stars at least 3 times the Einstein radius; their typical separations are $9.6, 5.2, 3.1$AU for $1, 0.3$ and $0.1M_\odot$ systems respectively. Indeed, a number of wide-separation planets will be included, whose masses mostly concentrate around $100M_\oplus$. 
We restrict the events whose time scales are larger than 0.1 days. We find their event rates are relatively small. For example, for the $1M_\odot$ and 0.3 $M_\odot$ systems, the event rates are both of the order of $1.8\times 10^{-3}$ per million stars per year (these should be compared with the numbers in Table \ref{tab:results30}, $3.9\times 10^{-2}$ and $1.4\times 10^{-2}$ per million stars per year). For the  $0.1 M_\odot$ host star, there are no objects beyond three Einstein radii. So the inclusion of planets on wide-separations does not change our results significantly.

Our main result (shown in Fig.~\ref{fig:results}) is that
the predicted time-scale is about a factor of $\sim 16$ shorter than the
typical short time-scale events seen by the MOA collaboration \citep{Sum+11} while the event rate is
smaller by a factor of $\sim 13$. It is important to emphasise that
even with this predicted low rate, the FFP population can still in principle be probed by observations,
provided that the cadence is sufficiently high. The cadence requirement is already satisfied
for some fields of MOA-II and OGLE-IV, and for most fields of the KMTNet that will have a cadence of 10 min to 15 min (see \citealt{Hen+14} for detailed predictions). A space satellite such as WFIRST or Euclid, with its superior photometric precision, will have sensitivities to Moon-mass
planets (\citealt{Pen+13, Spe+15}). Note, however, our study assumes perfect detection efficiency. This may not be such a bad assumption for KMTNet or WFIRST due to their high cadence. In any case, the observational results can be corrected for detection efficiency and then compared to our results. An additional simplification we made is that we ignored the finite source effect. Although this is reasonable for Jupiter mass FFPs, it is not a good assumption for planets below 1-10 $\mearth$, and their detectability will be reduced compared that for a point source (\citealt{BR96}). So one will need to take into account the finite source size effect and the sampling effect in order to compare our predictions and real data.

How might we resolve the apparent discrepancy between  core accretion theory and
observational data? One possibility may be that observations by \cite{Sum+11} have
over-estimated the number of FFPs, for example, due to
blending or red noise in the data \citep{Bac+15}. In this regard, observations by other teams of these short-duration events will be valuable. Even more importantly, unique mass determinations using a combination of the finite source size effect (\citealt{WM94, Gou94b, NW94,Y+04}) and satellite parallax (e.g., from Spitzer, Kepler, and WFIRST, \citealt{Refsdal1966,Gou94a, Hen+16,Zhu+16}) will give definitive results on the nature of the short-time-scale events. 

On the theoretical side, more studies are needed to account for a missing population of FFPs with masses comparable to that of Jupiter. For example, gravitational instability, a scenario not considered here, may produce a population of gas giants \citep{Bos06}. Indeed, near-IR observations of nearby young clusters show evidence for isolated Jupiter-mass objects (for an example, see \citealt{P+16}),
although whether they can account for the required mass  is unclear (\citealt{Bow+15}, see Fig.\,4 in \citealt{Hen+16}).  Furthermore, \cite{ILN13} only studied planets formed and ejected around single stars. It may be that planets formed in binary systems can be ejected more
easily \citep{SF15}.  Even in the single star scenario, the ratio of the predicted frequency of gas giants in eccentric orbits compared to  that in nearly circular orbits is 5 times lower than that found in radial velocity surveys (see \citealt{ILN13} for more discussions).  
The underestimation in their prediction may be because the secular perturbations between gas giants were not yet fully incorporated. This may in turn lead to an underestimation of the frequency of ejections. More efficient mechanisms for the formation of Jupiter-mass FFP’s or distant Neptune-mass planets also need to be further explored. Undoubtedly, a precise determination of the mass spectrum of FFPs will provide an important test of different theories of planet formation.


\section*{Acknowledgments}
We would like to thank Richard Long for a careful reading of the manuscript and all the participants at the 20th microlensing
(2016) workshop in Paris for a stimulating meeting. 
This work was supported by the Strategic Priority Research
Program ``The Emergence of Cosmological Structures'' of the Chinese Academy of Sciences
Grant No. XDB09000000, and by the National Natural Science Foundation of China (NSFC)
under grant number 11333003 and 11390372 (SM). Work by WZ was supported by NSF grant AST-15168.

\bibliographystyle{mn2e}
\bibliography{microlens}

\end{document}